\documentclass[final]{elsart}
\usepackage{graphics}
\usepackage{epsfig}
\usepackage{subfigure}
\usepackage{amssymb,latexsym} 
\usepackage{mathrsfs} 

\renewcommand{\cal}{\mathscr} 
\newcommand{\ea}[1]{\begin{eqnarray}} 
\newcommand{\eb}{\end{eqnarray}} 
\def\shiftdown#1{#1\llap{\lower.04ex\hbox{#1}}}

\journal{Physics Letters B}
\begin{document}

\begin{frontmatter} 
  \title{Experimental and theoretical evidences for an intermediate 
  $\sigma$-dressed dibaryon in the $NN$ interaction}
  \author[inp,tph]{V.I. Kukulin},
  \author[pit]{P. Grabmayr},
  \author[tph]{A. Faessler},
  \author[dub]{Kh.U. Abraamyan},
  \author[pit]{M. Bashkanov},
  \author[pit]{H. Clement},
  \author[pit]{T. Skorodko}, and
  \author[inp]{V.N. Pomerantsev}
\address[inp]{Institute of Nuclear Physics, 
  Moscow State University, %\\
  Vorobievy Gory, Moscow 119992, Russia}  
\address[tph]{Institut f\"ur Theoretische Physik, 
  Universit\"at T\"ubingen, %\\ 
  Auf der Morgenstelle 14, D-72076 T\"ubingen, Germany}
\address[pit]{Physikalisches Institut, Universit\"at T\"ubingen, %\\ 
  Auf der Morgenstelle 14, D-72076 T\"ubingen, Germany}
\address[dub]{Joint Institute for Nuclear Research, 141980, Dubna, Moscow region, 
Russia}     

\date{\today}

\bigskip

\hfill \parbox{0.5\textwidth}{This paper is dedicated to the memory 
of Lev Landau in occasion of his centenary}

\maketitle

\begin{abstract}
  Numerous theoretical and experimental arguments are presented in favor of
  the generation of intermediate $\sigma$-dressed dibaryon in $NN$
  interaction at intermediate and short distances. We argue that this
  intermediate dibaryon can be responsible for the strong intermediate-range
  attraction and the short-range repulsion in the $NN$ interaction, and also
  for the short-range correlations in nuclei.  The suggested mechanism for the
  $\sigma$-dressing of the dibaryon is identical to that which explains the
  Roper resonance structure, its dominant decay modes and its extraordinary
  low mass. A similar transformation mechanism from the glue to the scalar
  field was discovered in $J/\Psi$ decays. The new experimental data on
  2$\pi$-production in the scalar-isoscalar channel produced in $pn$- and
  $pd$-collisions and in particular the very recent data on $\gamma\gamma$
  correlations in $p$C and $d$C scattering in the GeV region seems to
  corroborate the existence of the $\sigma$-dressed dibaryon in two- and three
  nucleon interactions.
\end{abstract}


\end{frontmatter}

%%%%%%%%%%%%%%%%%%%%%%%%%%%%%%%%%%%%%%%%%%%%%%%%%%%%%%%%%%%%%%%%%%%%%%%%%%%%%%%
\section{\label{sec:intro} Introduction}

It is well known now that the conventional one-boson-exchange (OBE) concept of the
$NN$ interaction suggested long ago in the classical work by Yukawa describes very
well the peripheral part of $NN$ interaction at distances $r_{NN}\gtrsim 1.4$~fm. 
However this accepted mechanism meets quite serious problems on the fundamental
level in the description of phenomena depending on the short-range behavior of the
interaction when two nucleons are overlapping ~\cite{Plae,Ann05,Kaiser,Oset}.  To give examples,
we point out the difficulties with the extra-large short-range cutoff parameters
$\Lambda_\pi$ or $\Lambda_\rho$, with high values for $\omega NN$-coupling
constant $g_{\omega{NN}}$, and with the ratio for the tensor-to-vector $\rho{NN}$
coupling $\varkappa_\rho$. Because of these problems the short-range part of the
$NN$ interaction is treated in modern $NN$-potential models and in numerous
effective-field-theory (EFT) approaches mainly on a purely phenomenological
basis~\cite{Entem}. Very recently, however new serious problems~\cite{Kaiser,Oset} with the
basic scalar $NN$ force at {\em intermediate ranges} have arisen. Using completely
different approaches, a few groups~\cite{Kaiser,Oset} have revisited the scalar-isoscalar
part of 2$\pi$-exchange force which has previously been considered as basic
mechanism for the strong intermediate-range $NN$ attraction in conventional
OBE-models. Now this part of the interaction was treated more consistently than
before, and as a result, the correlated 2$\pi$-exchange has lead to a strong
short-range {\em repulsion} and only very moderate weak peripheral $NN$
attraction. The authors of Ref.~\cite{Kaiser} summarize their findings as: ``Contrary
to common belief these processes (i.e. the correlated 2$\pi$-exchange at
$r_{NN}\sim 1$~fm) lead to negligibly small and repulsive corrections to the $NN$
potential''.  With the above new findings (see especially refs.~\cite{Oset}), thus
the conventional $t$-channel 2$\pi$-meson exchange mechanism cannot provide the
strong intermediate-range attraction between two nucleons.

Furthermore, the predictions of the most accurate few-body calculations based on
conventional force models disagree in quite numerous cases with the precise
experimental data for the few-nucleon systems~\cite{Huber,Groep,Ruan,Maeda}. One of
the most indicative examples is the very strong disagreement between the results of
the recent NIKHEF~\cite{Groep} and JLab~\cite{Nicco} experiments on the $^3$He
high-energy disintegration into the three-nucleon channel, viz.  the $^3{\rm
He}(e,e'pp)$~\cite{Groep}  and the $^3{\rm He}(\gamma,pn)p$ reaction~\cite{Nicco}
at $E_\gamma=0.35 -1.55$~GeV and the only existing Laget's  model.

In order to avoid these serious problems at short and intermediate ranges, a few of
the present authors~\cite{JPhys01,IntJ02,Shikh04,Yaf05} have developed some time ago a new model for
this part of the $NN$ interaction.  The new approach assumes that  when two nucleons
are approaching very closely to each other for distances $r_{NN}\lesssim 1$~fm, their quark
cores overlap and a new intermediate state, the dressed dibaryon, appears. The
respective interaction model has been developed jointly by two groups (in Moscow and
T\"ubingen), initially as interesting and quite successful alternative concept to the
conventional $t$-channel meson-exchange mechanism at short ranges.  Although this
concept was a plausible {\em conjecture} it was able to explain consistently the character
of short-range $NN$ correlations. However recently very convincing new direct
experimental data appeared which, on the authors opinion,  can be interpreted as
direct evidence of appearance of the dressed dibaryon in the $NN$ system and thus the
generation of the dressed six-quark bag looked for in $NN$ system a long
time~\cite{Brod,Jaf,Brown}.  After a short description of the dibaryon concept for
the $NN$ interaction, we recapitulate in this letter all the new experimental data
and give their qualitative interpretation in terms of the dibaryon concept.

%%%%%%%%%%%%%%%%%%%%%%%%%%%%%%%%%%%%%%%%%%%%%%%%%%%%%%%%%%%%%%%%%%%%%%%%%%%%%%%
\section{\label{sec:concept}
 Concept of the dressed dibaryon in $NN$
  interaction}
There was a very high activity in 80ies around dibaryons and their experimental 
manifestation (see e.g. the comprehensive reviews~\cite{Strak,Locher}). However in all this
activity dibaryons (no matter, they are narrow or broad) were considered as some 
multi-quark {\em exotic mode}. In sharp contrast to this previous activity, we treat the
intermediate dibaryon as a basic carrier of short-range $NN$ interaction, i.e. as a
{\em regular d.o.f.} 

\begin{figure}[hpt]
\begin{center}
\epsfig{file=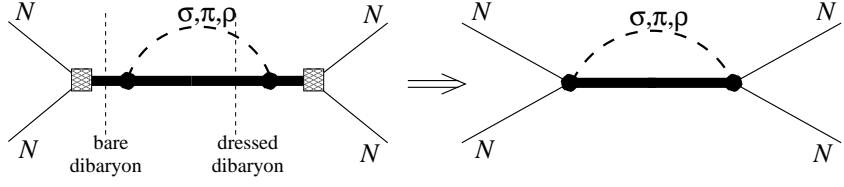,width=0.8\columnwidth}
\end{center}
\caption{\label{fig:graph}
  The graphs are illustrating the generation of the intermediate $s$-channel
  dibaryon dressed with $\sigma$, $\pi$ and $\rho$ meson fields. The
  contraction of the two bare dibaryon propagators in the left graph leads to
  the contracted graph shown in the right.}
\end{figure}

Thus, in our approach the $NN$ system is described formally as a two- (or multi-) component system
including at least two independent channels~\cite{JPhys01,IntJ02,JPhys3N}.  The external ($NN$)
channel describes the motion of two nucleons interacting with each other by a
conventional $t$-channel one- or two-meson exchange with the appropriate cut-offs at
short distances. However, when two nucleons are approaching each other, the system
changes to the inner -- dibaryon (DB) -- channel in which there are no individual
nucleons present at all and a new phase emerges which consists of the intermediate
dibaryon dressed with strong meson fields, e.g. $\sigma$, $\rho$ and $\pi$. Note,
that the main contribution is coming from the scalar $\sigma$-field surrunding
six-quark core in a space symmetric state.  Formally, it
may be interpreted in a way that the standard $t$-channel $\sigma$-exchange between two
nucleons at $r_{NN}\lesssim$~1~fm is replaced in our approach by the respective $s$-{\em channel}
$\sigma$-exchange associated with the intermediate dibaryon production (see
Fig.~\ref{fig:graph}). It is important to stress that all channels are defined in our
approach in a whole space.

%%%%%%%%%%%%%%%%%%%%%%%%%%%%%%% Fig.1 %%%%%%%%%%%%%%%%%%%%%%%%%%%%%%%%%%%%%%%%%
%%%%%%%%%%%%%%%%%%%%%%%%%%%%%%%%%%%%%%%%%%%%%%%%%%%%%%%%%%%%%%%%%%%%%%%%%%%%%%%

 In general, the model includes a Fock-column with different components: $NN$,
6$q+$$\sigma$, 6$q+$$\pi$, etc.~\cite{JPhys3N}. The main problem in this approach
is finding the coupling between the external $NN$ channel and the
internal $DB$ channels, i.e. the transition vertex $NN\to DB+\sigma$.\\
{\bf In the non-relativistic treatment}, we tested two different ways for this coupling:
fully microscopic~\cite{IntJ02} and semi-microscopic~\cite{JPhys3N}. Within microscopic
six-quark model the $\sigma$-dressing mechanism  was shown~\cite{JPhys01,IntJ02,Yaf05} to come from specific
mixed-symmetry six-quark configurations $|s^4p^2[42]_X[51]_{FS}\rangle$
dominating in the overlap region of two nucleons~\cite{Kusain,Stancu}. Thus, the
scalar $\sigma$-field can be easily generated in the transition
$|s^4p^2[42]_X[51]_{FS}\rangle\to |s^6[6]_X+\sigma\rangle$ in which two
$p$-shell quarks jump down to the $s$-shell with emission of two strongly
correlated $S$-wave pions. These two pions produce quite naturally a scalar
$\sigma$-meson~\cite{JPhys01,IntJ02} in the field of a six-quark core.

This treatment leads to a very simple and transparent model for interaction in
the external $NN$ channel:
\begin{equation}
\label{vtot}
V_{NN}=V_{OPE}+V_{TPE}+V_{NDN}+V_{orth}
\end{equation}
where $V_{OPE} (V_{TPE})$ are the peripheral \mbox{one-(two-)} pion exchange
interactions smoothly cutoff at $r\sim$~1~fm.  The intermediate- and
short-range term $V_{NDN}$ is a nonlocal potential of separable type coming
from intermediate $s$-channel dressed dibaryons (cf. Fig.~\ref{fig:graph}),
which in the simplest case at low and intermediate
($E_p\lesssim$~1~GeV) energies takes the form:
\begin{equation}
\label{vnqn}
V_{NDN}=\sum_{JLL'}\varphi_L^{J}({\bf r})\
                  \lambda_{LL'}^J(E)\,\varphi_{L'}^{\star J}({\bf r'})
\end{equation}
where $\varphi_L^J({\bf r})$ are the 2$s$ (or 2$d$) or 3$p$ (or 3$f$) h.o.
functions for the interaction in the $L^{th}$ partial wave.  The energy-dependent 
coupling constant $\lambda_{LL'}^J(E)$ is expressed in terms of the
loop integral (cf.~Fig.~\ref{fig:graph}) taking the form
\begin{equation}
\label{coupl}
\lambda_{LL'}^J(E)= \int_0^\infty\,d{\bf k}\,
\frac{B_L^J({\bf k},E)\cdot B_{L'}^{\star J}({\bf k},E)}
     {E-E_{DB}-k^2/2\overline{m}_\sigma}
\end{equation}
where $\overline{m}_\sigma$ is the reduced mass in the $6q+\sigma$ channel,
$k^2/2\overline{m}_\sigma$ is the kinetic energy of the $\sigma$-meson, $E_{DB}$
is the difference of the bare mass of the dibaryon and sum of nucleon masses.
$B_L^J$ and $B_{L'}^{J}$ are the vertex functions for the $DB+\sigma$ couplings
which can be calculated microscopically~\cite{IntJ02}. In the semi-microscopic variant of
the model the explicit consideration of multi-quark dynamics was replaced by
a simple parametrization of the vertex functions~\cite{JPhys3N}. The results of the
microscopic and semi-microscopic variants occured to be quite near to each other.

$V_{orth}$ is a pseudo-potential providing the orthogonality
condition between the excited 2$\hbar\omega$ and the non-excited 0$\hbar\omega$
6$q$-states expressed through the variables of the $NN$ channel. The potential in
the $S$-wave takes the form 
\mbox{$V_{orth}=\mu\,\,\langle{\bf r}|\varphi_0\rangle\langle\varphi_0|{\bf r}^\prime\rangle$}. 
Here $\varphi_0({\bf r})$  is the 0$s$ h.o. wave function and the constant $\mu$
should be taken positive and sufficiently large to eliminate the contribution of
the $s^6$ bag-like configuration from the initial $NN$ channel~\cite{IntJ02,Shikh04}.
Similar constructions for  $V_{orth}$ are found also for the $P$-waves.  It should
be noted, that in our approach the traditional mechanism for the short-range $NN$
repulsion induced by the vector (i.e. $\omega$- and $\rho$-) meson exchange is
replaced mainly by a nonlocal repulsion coming from the $V_{orth}$ term in
Eq.~(\ref{vtot}) and also from the nodal character of all $NN$ wave functions in
the low partial waves~\cite{JPhys01,Kusain,Stancu} with stationary inner nodes.%\footnote%

  Moreover, the inner node positions have been found to almost coincide with the
repulsive core radii in traditional $NN$-potential models. {The conventional
$t$-channel $\omega$- and $\rho$-exchanges  can also  be included into the
model with proper coupling constants but the main part of short-range $NN$
repulsion will come just from dibaryon mechanism. Thus, in our approach the
$\omega NN$ and $\rho NN$ coupling constants can be taken in full agreement with
$SU(3)$ prescription.} 

{\bf In a fully relativistic treatment}~\cite{Ann05,Shikh04} we start with a time ordered
two-point 
Green function for the transition chain $NN\to D(CC)\to NN$
\begin{eqnarray}
\label{g}
&&{\cal M}_{fi}=-\frac{i}{2!}\int d^4x_1d^4x_2d^4x_3d^4x_4\nonumber\\
&&~~\langle 4,3|T\{{\cal L}_{NND}(x_3,x_4)
\times{\cal L}_{NND}(x_1,x_2)\}|2,1\rangle 
\end{eqnarray}
where  $D(CC)$ means the confined (on color) dibaryon state and $T$ means a chronological ordering operator whereas the states
$|2,1\rangle$ and $\langle 4,3|$ correspond to the initial and final nucleons
with 4-momenta $p_2$,~$p_1$ and $p_4$,~$p_3$ respectively. The (nonlocal)
Lagrangian density is chosen as
\begin{eqnarray}
\label{l}
{\cal L}_{NND}(x_1,x_2)\,=\,\tilde N(x_2)\,\{\Psi(x_1,x_2)V_0(x_1-x_2)
                                                                    \nonumber\\
\,\,+\,\,\gamma_5\gamma^\mu\Psi_\mu(x_1,x_2)V_1(x_1-x_2)\}\,N(x_1)+h.c.
\end{eqnarray}
where function $\Psi(x_1,x_2)$ describes the dibaryon with spin $S=$~0 whereas
$\Psi_\mu(x_1,x_2)$ corresponds to the $S=$~1 dibaryon. In the two cases the
isospin $I$ is taken zero, $I=$ 0. For the $I=$ 1 dibaryon an evident replacement
$\Psi\to \vec\tau\vec\Psi$ should be done. After a lengthy calculation (for
details see Refs.~\cite{Ann05,Shikh04}) one gets a relativistic dibaryon-induced $NN$
potential which gives in the nonrelativistic reduction a separable potential
similar to Eqs.~(\ref{vnqn}) and~(\ref{coupl}).

In the simplest case, the above model~\cite{Yaf05,JPhys3N} has two components only, the
external $NN$ part and the inner $\sigma$-dressed dibaryon part. With this
two-component model we were able to fit excellently the lower partial wave $NN$
phase shifts up to 1000 MeV~\cite{Ann05,JPhys01,IntJ02}, using only a few basic parameters
(mass and radius of the intermediate dibaryon, and the cutoff parameters for
$V_{OPE}$ and $V_{TPE}$).   The quantitative estimates for the mass of the dressed
dibaryon still require the accurate consideration of the chiral symmetry
restoration effects and thus we postpone such cumbersome calculations to a future.
However we found that the $NN$ phase shifts can be nicely fitted until the energy
$E_{\rm Lab}\simeq 1$~GeV at the dibaryon mass $m_{DB}=2.2\div 2.3$~GeV (for
$^1S_0$ and $^3S_1- {}^3D_1$ partial waves).

Using two-component model one can calculate accurately
the weight~$W_D$ of the dressed dibaryon in the deuteron and also in the $NN$
scattering states.  For the deuteron we found $W_D\approx$~0.025~\cite{IntJ02,JPhys3N}.
Despite this rather small admixture of the dressed dibaryon component, it
provides (jointly with the OPE-tensor force and some very moderate peripheral
attraction coming from TPE) a sufficient intermediate-range attraction to fit
the $NN$ phase shifts up to 1~GeV and to bind the deuteron.

On the other hand, the qualitative picture of the intermediate dibaryon can also be
confronted with the earlier dibaryon models~\cite{Ping,Lomon} which have used the
different mechanisms for stabilization of intermediate six-quark bag (e.g.
delocalization of single-quark orbits~\cite{Ping} or the matching to external $NN$
channel on some surface~\cite{Lomon}. However, these previous models were operating
only with inner single-quark (and external $NN$) degrees of freedom without explicit
scalar field which plays a crucial role in our approach (see below) and is seen
clearly in experiments. The other alternative dibaryon model~\cite{Mulders} was 
built on a string-like picture, where the dibaryon was modeled as $4q+2q$ quark
clusters connected by a color string. However, the string in such models was in its
{\em ground state} regarding its radial excitation and there were also no meson
fields present. In terms of the color string between two quark clusters at its ends,
our picture corresponds to the generation of a 2$\hbar\omega$-excited string
configuration between two colored quark clusters in the initial state (because the
six-quark configuration $s^4p^2[42]_X$ includes two quanta of gluonic
excitation)~\cite{Ann05,IntJ02}.  Consequently, when the string deexcites with the
emission of two gluons the latter can transform their energy to the scalar
$\sigma$-field and eventually to two final pions. The energy transformation can be
easily understood if the $\sigma$-meson contains some glue component in form of
light glueball admixture.  This picture of the $\sigma$-generation from the string
deexcitation in the intermediate dibaryon is in nice agreement with recent
approaches~\cite{Kiss05,Kiss99} to the $\sigma$-meson emission in $pp$ collisions at
high energies. The authors suggested to estimate the one- and two $\sigma$-meson
emission from the intermediate string by assuming a hybrid nature of the light
scalar meson which has both glueball (0$^{++}$) and $\pi^0\pi^0$ ($\pi^+\pi^-$)
components. From this point of view, production of  the intermediate
$\sigma$-dressed dibaryon can be viewed as $s$-channel version of the
$\sigma$-emission mechanism (proposed by Kisslinger \etal for Pomeron exchange at
high energies~\cite{Kiss99}) in case of low and intermediate energies.

\section{\label{sec:appl}
Application to the Roper resonance structure and decay}

The above dibaryon model combined with the specific mechanism for the strong
scalar field generation in the quark bag gives unique predictions for several
hadronic processes and for many properties of the few-body systems.  In
particular, the non-conventional picture for the strong scalar field
generation described above was confirmed very nicely in recent studies of the
Roper-resonance structure~\cite{Dillig} and its decay~\cite{Skorod08,Bashk07,Skorod05} and also in other
new experiments with high-energy deuterons~\cite{Ab06}. In fact, the quark structure of the Roper
resonance (corresponding also to the 2$\hbar\omega$ monopole excitation of the
initial nucleon) should include the excited 3$q$-configurations
$|(0s)(1p)^2[3]_X+(0s)^2(2s)[3]_X)\rangle$, i.e. this Roper state includes
also two $p$-shell quarks (or a single quark in the $2s$-excited state).
Hence quite similar to our dibaryon, one has for the Roper-resonance a
generation mechanism for the scalar field:
$|sp^2[3]_X\rangle\to|s^3[3]_X+\sigma\rangle$.  This means that the dominating
configuration for the Roper resonance might be $|s^3[3]_X+\sigma\rangle$, i.e.
the $\sigma$-dressed three-quark bag in a space symmetrical 3$q$-state $|s^3\rangle$ which
is equivalent -- in the 3$q$-dynamics -- to our intermediate dibaryon
production mechanism in the 6$q$-dynamics.  Actually, the large admixture of
such a $\sigma$-field configuration to the 3$q$-state $(0s)^3$ was found both
in theoretical~\cite{Dillig} and experimental~\cite{Skorod08,Bashk07,Skorod05} studies. Moreover,
two independent studies~\cite{Nary,Kiss97}, both done on basis of QCD sum
rules, have found that the Roper resonance state has a large admixture for the
$\sigma$-field while the nucleon ground state has a negligible amount of
$\sigma$-meson admixture.

These conclusions have been well confirmed in a recent dedicated
experiment~\cite{Skorod08,Skorod05} where the $\sigma$-meson channel was found to be
strongly dominating in the Roper-resonance decay with a branching ratio
$\Gamma(R\to N+\sigma)/\Gamma(R\to \Delta+\pi)$ of $\sim$~4:1. Thus, the high
admixture of the scalar field in the Roper wave function should be one of the
main reasons for the strong shift downwards of the Roper-resonance mass to
anomalously low mass value $m_R\approx~1370~MeV$~\cite{Skorod05}). A very similar
shift downwards of the $\sigma$-dressed dibaryon mass enhances significantly
the $NN$ intermediate-range attraction via the mechanism shown in
Fig.~\ref{fig:graph}.

%%%%%%%%%%%%%%%%%%%%%%%%%%%%%%%%%%%%%%%%%%%%%%%%%%%%%%%%%%%%%%%%%%%%%%%%%%%%%%%
\section{\label{sec:symres}
The partial chiral symmetry restoration in dibaryon dynamics}

The dressed dibaryon concept is very tightly related to the idea about partial chiral
symmetry restoration (CSR) in dense hadronic matter ~\cite{Hatsuda}. As is well known, the
$\sigma$-meson mass in such a symmetry restoration process decreases drastically and
this leads to a strong stabilization of the most symmetrical $6q$-configuration
$|s^6[6]_X\rangle$ embedded into this scalar field.  Simultaneously the mass of the
intermediate dressed dibaryon comes also down and as a result of all these highly
non-linear effects there appears a strong effective attractive force between two nucleons at
intermediate distances $r_{NN}\sim 1$~fm. {In the conventional quantum-mechanical
language this strong intermediate-range attraction in $NN$ channel can be interpreted
in terms of coupled-channel phenomena, i.e. as a result of strong coupling between
the external $NN$ channel and inner dibaryon channel with symmetry $|s^6[6]_X\rangle$
dressed with $\sigma$-field.}  This picture should be compared with result of pure
$6q$-model~\cite{Stancu}, i.e. that without any scalar fields where $qq$ force
(originating from the Goldstone-boson-exchange (GBE) mechanism) are fitted very
nicely to reproduce the spectra of excited baryons (in the normal and strange
sectors)~\cite{StGloz}.  However, the effective two-nucleon force resulting from the GBE
$qq$ force was found~\cite{Stancu} to be {\em purely repulsive} and strong in the nucleon
overlap region. Thus, the $qq$ force which fits very well the baryon spectra (octets,
decuplets et.) should
be supplemented in the $NN$ system with an additional strong scalar force (either on
the $qq$- or directly on the $NN$ level) to fit the deuteron properties and the $NN$
scattering data.  We conjecture, the dibaryon mechanism considered here is very well
suited to be origin for this strong scalar force.

Furthermore, in combination with recent ideas about chiral symmetry 
restoration~\cite{Hatsuda,Gloz} the dibaryon model predicts the strong enhancement for the
$\sigma$-meson production in $NN$ collisions in the energy region $E_p^{lab}\approx$
1.1 - 1.2~GeV. This property follows straightforwardly from untiing the $\sigma$ loop
of the dressed bag propagator (cf. Fig.~\ref{fig:graph}) at energies above the
threshold of $\sigma$-meson production. 

As a clear signal for such a chiral symmetry restoration (CSR) effect one considers
usually parity doublets in the baryon spectra at high excitation~\cite{Gloz}.  The
first (rather approximate) parity doublet is suggested~\cite{Gl07} to be the Roper
$P_{11}$(1440) ($J^P=1/2^+$) and $S_{11}$(1535) ($J^P=1/2^-$) isobars. Thus, from
this point of view, the 2$\hbar\omega$-excitation of the dressed dibaryon is very
similar to the 2$\hbar\omega$-excitation of the Roper-resonance. So, the scalar-isoscalar
enhancement near the 2$\pi$-threshold, associated with the intermediate
$\sigma$-meson in case of CSR in the excited multiquark bag, should be seen in
experiments on two-pion ($\pi^0\pi^0$ or $(\pi^+\pi^-)_{00}$) production in $NN$,
$Nd$ etc.  collisions.  Fortunately, such experiments have been done numerously in
60ies~\cite{Booth} and the authors found very clear and strong enhancement in the
2$\pi$-production near the thresholds in $p+d\to {}^3He+(\pi\pi)_{00}$, $d+d\to
{}^4He+(\pi\pi)_{00}$ reactions. These experimental results shown as missing mass
distributions are well known nowadays as the famous ABC-puzzle~\cite{Booth,Bashk06}.

This (partial) symmetry restoration effects have been confirmed by two independent
groups of authors: first, this restoration was found~\cite{Hatsuda} in large hadronic
systems at high temperature or density, and second, this effect was demonstrated to
happen even in a single hadron at high excitations~\cite{Gloz,Gl07}. There are also
quite general arguments in favor of the $\sigma$-mass renormalization~\cite{Hatsuda,Gloz} at
sufficiently high excitation energy in both cases mentioned above.

%%%%%%%%%%%%%%%%%%%%%%%%%%%%%%%%%%%%%%%%%%%%%%%%%%%%%%%%%%%%%%%%%%%%%%%%%%%%%%%
\section{\label{sec:exp}
New experimental evidence for the $\sigma$-dressed dibaryon}

Very recently two types of new experimental data appeared~\cite{Bashk07,Skorod05,Ab06,Bashk06}
which have given the direct and unambiguous evidence for the intermediate
$\sigma$-dressed dibaryon production with strongly renormalized $\sigma$-meson
mass. The first type of experiments~\cite{Bashk07,Bashk06} is in essence an improvement
of the old classical ABC experiments with modern exclusive setting and
detailed measurements of energy and angular correlations of two emitted pions
in the reactions $p+d\to {}^3{\rm He} + \pi^0\pi^0 \mbox{ (or } \pi^+\pi^-)$
and $p+n\to d+\pi^0\pi^0$ at incident proton energies ($E_p^{lab}\sim
1.1-1.2$~GeV) specific for the manifestation of the ABC phenomenon.
%%%%%%%%%%%%%%%%%%%%%%%%%    Fig.2    %%%%%%%%%%%%%%%%%%%%%%%%%%%%%%%%%%%%%%%%%
\begin{figure}[hptb]
\begin{center}
\epsfig{file=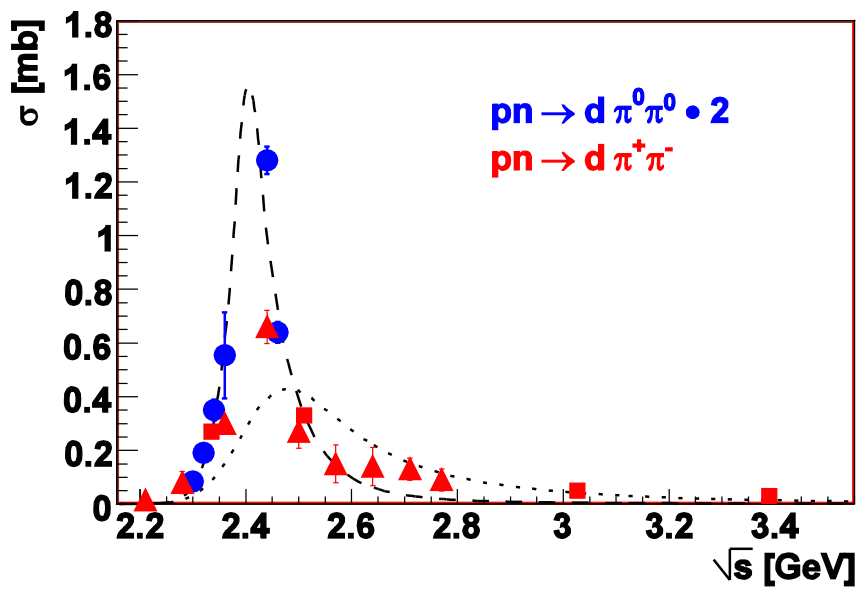,width=0.9\columnwidth}
\end{center}
\caption{\label{fig:abc}
  The excitation functions for the reactions
  $pn\to{d}\pi^\circ\pi^\circ$ and $pn\to{d}\pi^+\pi^-$.
  Predictions of the conventional $\Delta\Delta$ model\cite{BarNir} are shown with 
  dotted line while the long dashed line corresponds to the fit with the
  near-threshold $\Delta\Delta$-bound state  (for detail, see refs.~\cite{Bashk07}).
}
\end{figure}
%%%%%%%%%%%%%%%%%%%%%%%%%%%%%%%%%%%%%%%%%%%%%%%%%%%%%%%%%%%%%%%%%%%%%%%%%%%%%%%

The new experimental data of the CELSIUS-WASA collaboration~\cite{Skorod08,Bashk07,Skorod05,Bashk06}
together with those from earlier measurements have demonstrated
(Fig.~\ref{fig:abc}) that the rather narrow and strong peak observed in the
$p+n\to d+(\pi\pi)_{00}$ cross section cannot be explained by the conventional
$\Delta\Delta$ model~\cite{BarNir} (dotted line in Fig.~\ref{fig:abc}), and the
data require a near threshold $\Delta\Delta$ {\em bound state} (long dashed line).
In fact, a few $\Delta\Delta$ bound states have been predicted in some recent
six-quark calculations~\cite{Ping,Valcarce,Yuan}. However, in all these calculations
the r.m.s. radius $r_{\Delta\Delta}$ of matter in such $\Delta\Delta$ bound
states was found to not exceed 0.9~fm~\cite{Valcarce}, or to range even at $0.72 -
0.82$~fm~\cite{Yuan}.  Thus, two deltas in such bound states are strongly
overlapping and therefore the assumed $\Delta\Delta$ bound states are
nothing else but the intermediate dibaryon components. Another important result of
these experiments is that the authors found (similarly to the ABC-group) a strong
enhancement in $(\pi\pi)_{00}$-production near threshold at $M_{\pi\pi}\sim
320-330$~MeV with clear $S$-wave $\pi\pi$-correlation. Hence, in this case,
the ABC puzzle can be considered as an indicator for partial chiral symmetry
restoration in the excited six-quark system generated from collision of two
nucleons at energies $E_p^{lab}\sim 1.1-1.2$~GeV.  In such a specific
generation process, all the kinetic energy of the two-nucleon relative motion
($\approx$0.6~GeV) transforms into a 2$\hbar\omega$ string excitation near the
$\Delta\Delta$ threshold with subsequent $\sigma$-meson emission.

%%%%%%%%%%%%%%%%%%%%%%%%%    Fig.3    %%%%%%%%%%%%%%%%%%%%%%%%%%%%%%%%%%%%%%%%%
\begin{figure}[hptb]
\begin{center}
\epsfig{file=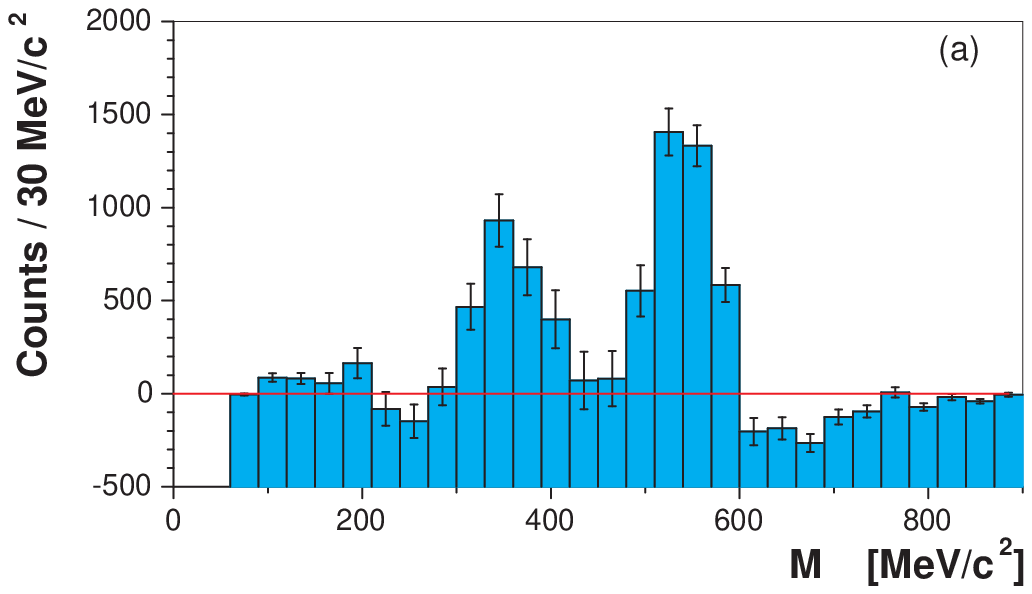,width=0.9\columnwidth}
\epsfig{file=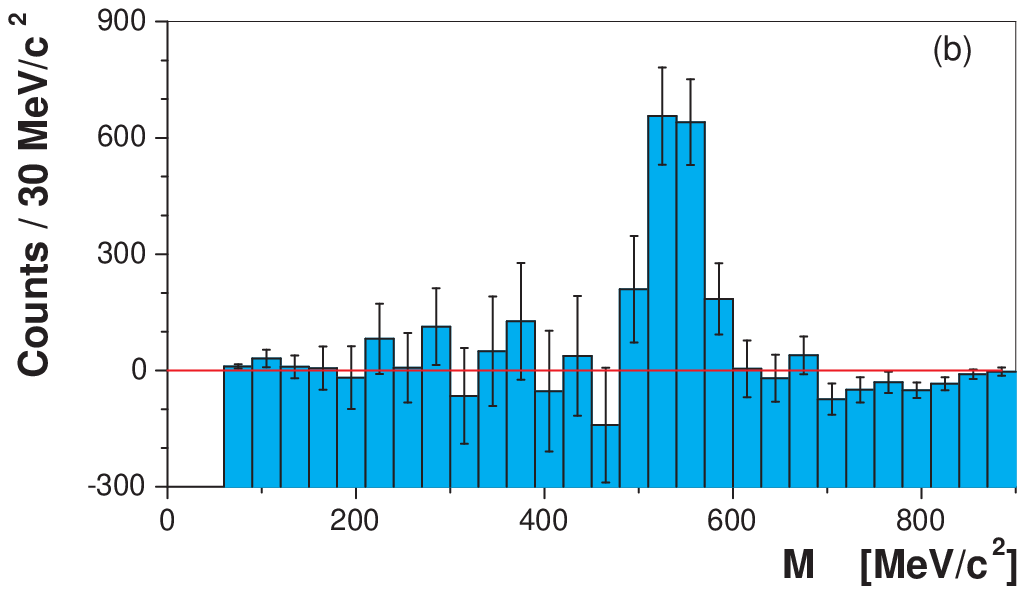,width=0.9\columnwidth}
\end{center}
\caption{\label{fig:imgg}
Invariant mass distributions for pairs of $\gamma$-quanta (a) in the
  reaction $d+{\rm C}\to\gamma+\gamma+X$ at a momentum of incident deuterons
  2.75~GeV/c per nucleon and (b) in the reaction $p+{\rm C}\to\gamma+\gamma+X$
  at momentum of 5.5~GeV/c, after background subtraction~\cite{Ab06}.}
\end{figure}
%%%%%%%%%%%%%%%%%%%%%%%%%%%%%%%%%%%%%%%%%%%%%%%%%%%%%%%%%%%%%%%%%%%%%%%%%%%%%%%

Another very clear signal for the $\sigma$-meson associated with the
renormalized-mass $\sigma$-dressed dibaryon in the deuteron has been found in
experiments $p+{\rm C}\to X+\gamma\gamma$ and $d+{\rm C}\to X+\gamma\gamma$ at
incident energies 2.5 - 5~GeV/N at Dubna~\cite{Ab06}. In the $d+$C experiment (see
Fig.~\ref{fig:imgg}a) the authors have found three clear peaks in the
$\gamma\gamma$-mass ($M_{\gamma\gamma}$) distribution: the low-energy peak for
$M_{\gamma\gamma}\simeq 140$~MeV associated with the $\pi^0$-decay mode was
suppressed in the analysis by cuts (and so not seen in Fig.~3), a high-energy peak
at $M_{\gamma\gamma}\sim~540$~MeV associated with the $\eta$-production and a
strong enhancement at $M_{\gamma\gamma}\sim~355$~MeV exhibiting a width
$\Gamma_{\gamma\gamma}\sim 49$~MeV; the latter could not be interpreted by the
authors~\cite{Ab06} in terms of well known mesons despite the comprehensive and
detailed modeling.

A similar peak at $M_{\gamma\gamma}\sim 300-320$~MeV has been found by the
CELSIUS-WASA collaboration~\cite{Ba04} in $pp\to pp\gamma\gamma$ experiments. This
peak has been interpreted by the authors as a signal of the intermediate
$\sigma$-meson with renormalized mass in the process $pp\to pp\sigma\to
pp\gamma\gamma$.  However, the authors~\cite{Ba04} have interpreted two gammas
emitted from the reaction as emerging from the intermediate $\pi\pi$ bremsstrahlung.
Contrary to this, the new data of the Dubna group have a quite unambiguous
interpretation just through the $\sigma$-dressed dibaryon component in the incident
deuteron because the signal is seen very clearly in $d+$C collisions and is not seen
in $p+$C collisions (cf. Fig~\ref{fig:imgg}a and~\ref{fig:imgg}b). In the first case
the $\sigma$-cloud of the dressed dibaryon component in the incident deuteron is
picked-off of six-quark core through the interaction with the carbon target producing
a $\gamma\gamma$ signal ($\sigma \to \gamma\gamma$) with a well known branching ratio
$\Gamma_\sigma(\gamma\gamma)/\Gamma_\sigma(\pi\pi)\sim~10^{-5}-10^{-6}$. Our first
estimates for the $\gamma\gamma$-yield in the $d+$C process are in an approximate
agreement with the measured number of the $\gamma\gamma$ events.

The suggested common mechanism for the generation of a light scalar field in the
Roper resonance and in dressed dibaryon dynamics was observed actually also in the
$J/\Psi$ decays studied by the BES collaboration~\cite{Ablikim}, the Crystal Barrel
group~\cite{Anis1,Anis2} and by the Fermilab E760 experiment~\cite{Armstrong}.  Here the
$J/\Psi$-mesons are produced often in the states with radial excitations of
$c\bar{c}$ string.  The authors~\cite{Anis1,Anis2} have found the decay of the excited
$0^{++}$ states (after $\gamma$-emission from excited $J/\Psi$ states) to proceed
with dominating emission of double $\sigma$-mesons. These specific transitions of
$c\bar{c}$ excited strings give a good {\em independent evidence} for the
discussed mechanism of the scalar field generation and thus in a way for
the $\sigma$-dressed intermediate dibaryon in the $NN$ system.

Summarizing all these completely independent experimental findings  one can
conclude that there should be a common mechanism for the enhanced $\sigma$-meson
emission from the Roper resonance decay, from  the 2$\hbar\omega$-excited intermediate
dibaryons, from high-energy $dC$ collisions, and also from deexcitation of
$c\bar{c}$ strings. This common mechanism can be further confirmed  with many
theoretical arguments presented here about the $\sigma$-meson hybrid nature and
chiral symmetry restoration in excited hadrons. So that the concept of
$\sigma$-field dressing of the intermediate dibaryon in $NN$ system (or
alternatively $2\hbar\omega$ string deexcitation mechanism into an enhanced scalar
$\sigma$-field) looks to be quite convincing one.  Simultaneously, it seems prove
the generation of the intermediate $6q$ bag in deuteron and $NN$ system predicted
long  ago~\cite{Brod,Jaf,Brown} on the basis of general arguments of quark model
and some experimental data. Thus, the above dibaryon concept seems to open a
door to the QCD-based description for nuclear phenomena.

%%%%%%%%%%%%%%%%%%%%%%%%%%%%%%%%%%%%%%%%%%%%%%%%%%%%%%%%%%%%%%%%%%%%%%%%%%%%%%%
{\bf Acknowledgments.} The authors appreciate very much the numerous discussions with
many colleagues from various research groups. Special thanks go to Dr. I.~Obukhovsky
who collaborated with us in many works and to Prof. S.~Moszkowski for many
stimulating discussions. V.I.K is deeply grateful to Prof. S.~Brodsky for his strong
encouragement of this direction of our work and useful comments. A partial financial
support from the DFG grant No.~436~RUS113/790 and GRK 683, and from BMBF 261 and the
RFFI grant No.~08-02-91959 is appreciated.

%%%%%%%%%%%%%%%%%%%%%%%%%%%%%%%%%%%%%%%%%%%%%%%%%%%%%%%%%%%%%%%%%%%%%%%%%%%%%%%

\end{document}